\documentclass[letterpaper,journal]{IEEEtran}

\usepackage{algorithmic}
\usepackage{algorithm}
\usepackage{array}
\usepackage{textcomp}
\usepackage{stfloats}
\usepackage{verbatim}
\usepackage{balance}

\usepackage{graphicx}
\usepackage[caption=false,font=scriptsize,labelfont=sf,textfont=sf]{subfig}

\usepackage{microtype}
\usepackage{url}
\usepackage{combelow}
\newcommand{\ii}[1]{{\footnotesize \textcolor{gray}{#1}}}

\usepackage{latexsym}
\usepackage[cmex10]{amsmath}
\usepackage{amssymb}
\usepackage{wasysym}
\usepackage{cancel}
\usepackage{bm}

\usepackage{booktabs}
\usepackage{multirow}
\usepackage{tabularx}
\newcommand{\mytable}{
	\centering
	\renewcommand{\arraystretch}{1.2}
}
\newcolumntype{C}{>{\centering\arraybackslash}X}
\newcolumntype{L}{>{\raggedright\arraybackslash}X}
\newcolumntype{R}{>{\raggedleft\arraybackslash}X}
\newcolumntype{P}[1]{>{\raggedright\arraybackslash}p{#1}}
\usepackage{siunitx}
\usepackage{etoolbox}                           
\newcommand{\ubold}{\fontseries{b}\selectfont}  
\robustify\ubold                                

\usepackage{cite}
\newcommand{\citet}[2]{#1~\cite{#2}}

\usepackage[prependcaption,textsize=scriptsize]{todonotes}
\definecolor{mycolor}{HTML}{FF6600}

\begin{document}

\title{Should Top-Down Clustering Affect Boundaries \\in Unsupervised Word Discovery?}

\author{
    \IEEEauthorblockN{
        Simon Malan, Benjamin van Niekerk, Herman Kamper
    }\\
    \IEEEauthorblockA{
        \textit{Electrical and Electronic Engineering,         Stellenbosch University, South Africa}\\
        {
            \small
            24227013@sun.ac.za, benjamin.l.van.niekerk@gmail.com, kamperh@sun.ac.za
        }
    }
}

\IEEEpeerreviewmaketitle
\maketitle

\begin{abstract}
We investigate the problem of segmenting unlabeled speech into word-like units and clustering these to create a lexicon.
Prior work can be categorized into two frameworks.
Bottom-up methods first determine boundaries and then cluster the fixed segmented words into a lexicon.
In contrast, top-down methods incorporate information from the clustered words to inform boundary selection.
However, it is unclear whether top-down information is necessary to improve segmentation.
To explore this, we look at two similar approaches that differ in whether top-down clustering informs boundary selection.
Our simple bottom-up strategy predicts word boundaries using the dissimilarity between adjacent self-supervised features, then clusters the resulting segments to construct a lexicon.
Our top-down
system is an updated version of the ES-KMeans dynamic programming method that iteratively uses $K$-means to update its boundaries.
On the five-language ZeroSpeech benchmarks, both approaches achieve comparable state-of-the-art results, with the
bottom-up system being nearly five times faster.
Through detailed analyses, we show that the top-down influence of ES-KMeans can be beneficial (depending on factors like the candidate boundaries), but in many cases the simple bottom-up method performs just as well.
For both methods, we show that the clustering step is a limiting factor. 
Therefore, we recommend that future work focus on improved clustering techniques and learning more discriminative word-like representations.
Project code repository: \url{https://github.com/s-malan/prom-seg-clus}.
\end{abstract}

\begin{IEEEkeywords}
word segmentation, lexicon learning, zero-resource speech processing, unsupervised learning.
\end{IEEEkeywords}

\section{Introduction}

\IEEEPARstart{U}{nsupervised} word segmentation aims to identify word boundaries in unlabeled speech.
This is challenging since speech is a continuous stream
without obvious silences between words~\cite{okko_early_lan_acq}. 
Constructing a lexicon poses another challenge, as no two speakers are identical---even individual speakers show a lot of variation in their speech.
Human infants navigate these challenges remarkably well, demonstrating
word discrimination and recognition capabilities within their first year~\cite{elika_6_9_word_meaning, native_lan_acq}. 
Solving the problem of segmenting and clustering speech could provide a way to improve
our understanding of human language acquisition~\cite{emmanuel_la_reverse_eng}.
It could also advance the development of low-resource speech technologies~\cite{aaurent_asr_low_resource}.

Early word discovery methods
employed direct pattern matching, usually using dynamic time-warping~\cite{park_seg_dtw}, to find matching segments in pairs of utterances. 
Although these methods have progressed~\cite{okko2020probDTW,dusted}, they fail to discover patterns that cover all the speech audio. 
In this paper, we focus on full-coverage systems that provide a full tokenization of the input speech into word-like units.

Several full-coverage strategies have been proposed~\cite{lee+etal_tacl15,algayres+etal_tacl22,okuda2022double}.
These can be roughly divided into two frameworks.
The first covers bottom-up methods that predict word boundaries and then build a lexicon based on the hypothesized segments~\cite{okko_syll_kmeans, bhati_se_ae, herman_dpdp, herman_dpdp_hu}.
Here, the lexicon does not affect the discovered word boundaries.
On the other hand, the top-down framework uses higher-level information to inform boundary selection~\cite{herman_bes_gmm, bhati_scpc, cuervo_macpc}.
The older embedded segmental $K$-means (ES-KMeans) method is an example~\cite{herman_eskmeans}.
It employs an iterative segmenting and clustering scheme, where each iteration selects the current best boundary hypothesis based on how well it matches the $K$-means clustering model.
This method employs dynamic programming and remains competitive~\cite{herman_dpdp_hu}, despite utilizing older speech features and boundary constraints.

In this paper, we investigate whether top-down information is beneficial when determining the
segmentation.
To do this, we look at two representative full-coverage methods.
The two methods are similar, but differ in whether or not top-down clustering can inform bottom-up discovered boundaries.
Inspired by Pasad et al.~\cite{ankita_tti}, 
our bottom-up system finds word boundaries through a lightweight method measuring dissimilarity between adjacent self-supervised features.
An explicit lexicon is constructed by clustering the discovered word segments using $K$-means.
Our top-down approach is an updated version of ES-KMeans~\cite{herman_eskmeans}.
After our updates, ES-KMeans closely resembles our bottom-up method. 
Both consist of segmentation and clustering steps and both model whole words directly without intermediate subword representations~\cite{herman_dpdp}.
The difference is that ES-KMeans iteratively refines its boundaries using top-down information from clustering, while our bottom-up approach fixes the boundaries before learning a lexicon.

We compare our methods to state-of-the-art approaches on the English, French, Mandarin, German, and Wolof evaluations from Track 2 of the ZeroSpeech Challenge~\cite{ewac_zrc}.
On this benchmark, both of our methods achieve some of the best results across the five languages, with the
bottom-up method being almost five times faster than its top-down counterpart.

This paper extends the conference paper~\cite{malan2025}, where we focused on the bottom-up method but also gave some results for ES-KMeans.
Here, we expand our comparison beyond the ZeroSpeech benchmarks to determine if top-down influence can be justified given the additional computational cost.
Our work here
includes investigating what types of boundaries and units are discovered by the methods, and how the candidate boundaries affect top-down influence.
We make the following contributions
(1)~We introduce two similar methods, the first follows a simple bottom-up and the other a top-down methodology.
(2)~We show that the simple bottom-up method delivers comparable state-of-the-art results to the top-down approach while being much faster.
(3)~We investigate the differences between the two methodologies and show that top-down influence can be beneficial if the candidate boundary set requires refinement. However, in many cases, the simple bottom-up method is good enough.
(4)~We show that our lexicon-building step limits how well both our methods perform.

Taken together, our work leads to concrete recommendations for the ZeroSpeech community.
Speech representations have seen large improvements through major efforts over the last decades~\cite{livescu_ssl_review}, but they are still far from perfect when used to represent whole word-like segments.
There has been far less work on different clustering methods for lexicon building, and tailored clustering methods might be needed~\cite{kamper_slt2014}.
Candidate boundary selection has also seen major improvements~\cite{ankita_tti}, but we show that the best candidate approach is influenced by whether a bottom-up or top-down framework is followed.

\section{Background}
\label{sec:background}

In this section, we discuss previous methods for full-coverage unsupervised word discovery.
We can classify these methods as either bottom-up or top-down.

Bottom-up approaches determine word boundaries based on some local similarity metric.
The idea is that features within a word should remain relatively consistent, with larger changes across word boundaries~\cite{tenbosch2007, bhati_scpc, cuervo_macpc, ankita_tti}.
By detecting these changes, input speech can be divided into word-like segments.
Finally, the discovered segments are clustered to construct a lexicon.
In contrast to top-down methods, the word boundaries are fixed and not influenced by the clustering step.

Early work relied on signal processing methods to segment speech into word-like units.
In~\cite{okko_syll_kmeans}, syllables serve as the basic building blocks of these units.
First, syllable boundaries are detected by analyzing patterns of stress and sonority in the input speech.
Next, syllable segments are clustered and tokenized by labeling each segment with its cluster assignment.
Finally, repeating patterns in the discrete units are extracted to model multisyllabic words.
Similarly, Bhati et al.~\cite{bhati_se_ae} use recurring cluster assignments to identify word-like segments, placing boundaries at frames with low similarity to their preceding neighbors.
To measure frame similarity, they propose a self-expressing autoencoder designed to extract features that capture phonetic content.

More recently, duration-penalized dynamic programming (DPDP)~\cite{herman_dpdp} follows a bottom-up approach that encodes input speech as a sequence of discrete phone-like units before applying a symbolic segmentation step.
The original method stopped at segmentation, but~\cite{herman_dpdp_hu} extended this idea by clustering averaged HuBERT features from each segment to construct a lexicon.
Unlike these methods, our bottom-up approach 
models words directly, without requiring intermediate phone or syllable representations.

Our proposed bottom-up approach (Section~\ref{sec:psc}) is underpinned by the work of Pasad et al.~\cite{ankita_tti}. They
simply place boundaries at peaks in the distances between adjacent self-supervised features.
The method requires no training and is effective for word segmentation.
This demonstrates that modern self-supervised representations encode information about word boundaries without being explicitly trained for segmentation.
Building on this idea, we propose an additional clustering step to construct a lexicon from the hypothesized word segments.

In contrast to the bottom-up approaches described above, top-down systems leverage higher-level information to guide boundary selection.
This framework
argues that jointly training boundary detection with other model components benefits word discovery.

Two
examples are the contrastive predictive coding models proposed in 
\cite{bhati_scpc} and~\cite{cuervo_macpc}.
Using a contrastive loss, these models jointly learn hierarchical representations of speech at both the frame and segment levels.
Subsequently, Bhati et al.~\cite{bhati_scpc} identify word boundaries by detecting peaks in the dissimilarity curve between adjacent segments; no lexicon is produced.
This is similar to the bottom-up strategies of~\cite{bhati_se_ae} and~\cite{ankita_tti} but with the added influence of top-down information.

Two earlier top-down systems perform word discovery by iteratively refining an initial set of candidate boundaries~\cite{herman_bes_gmm,herman_eskmeans}.
Candidates are first generated through an unsupervised syllable segmentation method~\cite{okko_sylseg}. 
Leveraging top-down information from a clustering model, the idea is to select the syllable boundaries that align with words, while discarding the rest.
The models then
iteratively cluster and re-segment the utterances using top-down information from the clustering model.
The first system~\cite{herman_bes_gmm} 
clusters pre-trained acoustic word embeddings with a Gaussian mixture model and uses Gibbs sampling~\cite{resnik_gibbs} to jointly infer word boundaries and cluster assignments.
Building on this idea, the second method, embedding segmental $K$-means (ES-KMeans)~\cite{herman_eskmeans}, approximates full Bayesian inference using hard clustering and segmentation.
More details on ES-KMeans are given in Section~\ref{sec:eskm}.

Given the success of bottom-up methods, how important is top-down information?
To explore this question, we compare a bottom-up model inspired by Pasad et al.~\cite{ankita_tti} with a revised implementation of ES-KMeans.
We update ES-KMeans to use modern features and better candidate boundaries.
The result is a method that looks very similar to our bottom-up approach, except that top-down clustering affects boundary selection. As a result, we can isolate the effects of top-down information.
We start by describing our bottom-up method.

\section{Bottom-Up Word Discovery: Prominence-Based Boundaries with Clustering}
\label{sec:psc}

Our bottom-up full-coverage unsupervised word discovery system consists of two components. 
First, we determine word boundaries using a prominence-based approach. 
Second, we cluster the predicted word-like units to build a lexicon.

\begin{figure}[!t]
    \centerline{\includegraphics[width=\linewidth]{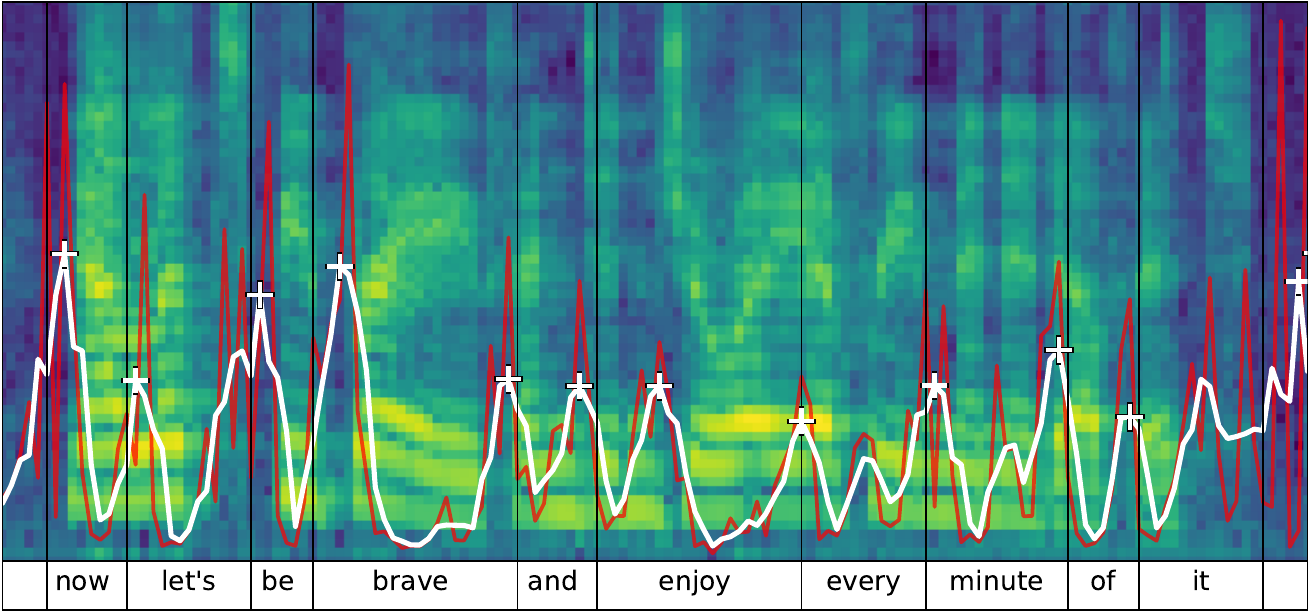}}
    \caption{An example of word boundaries from the prominence-based approach of~\cite{ankita_tti}. The red (dark) line is the dissimilarity curve between adjacent frames, which is smoothed to produce the white line. The crosses are the predicted boundaries. The black vertical lines are the ground-truth boundaries.}
    \label{fig:tti}
\end{figure}

For word boundary detection, we follow the method of Pasad et al.~\cite{ankita_tti}.
Speech utterances are first encoded by extracting features from an intermediate HuBERT~\cite{wei_hubert} layer, which we denote as $\mathbf{y}_{1:T}=\mathbf{y}_{1},\mathbf{y}_{2},...,\mathbf{y}_{T}$.
To predict word boundaries, the dissimilarity between neighboring frames $(\mathbf{y}_{t+1},\mathbf{y}_{t})$ is calculated using the cosine distance. 
Then, a moving average function is applied to the dissimilarity curve.
For this step, 
the HuBERT features are 
mean and variance normalized beforehand.
As illustrated in Fig.~\ref{fig:tti}, peaks on the smoothed dissimilarity curve are marked as word boundaries when the dissimilarity at frame $t$ is greater than some prominence threshold relative to the dissimilarity of surrounding frames. 
Using prominence instead of a hard dissimilarity threshold filters out some noisier peaks, leading to better word boundaries.
The underlying principle is that
frames within the same word are close to each other, while frames at word boundaries are further away from each other. 
The features are therefore crucial. 
This is why we use HuBERT, which attempts to encode phonetic content while removing speaker-specific information~\cite{wei_hubert, bshall_s_vs_d_units}.

\begin{figure}[!t]
    \centerline{\includegraphics[width=\linewidth]{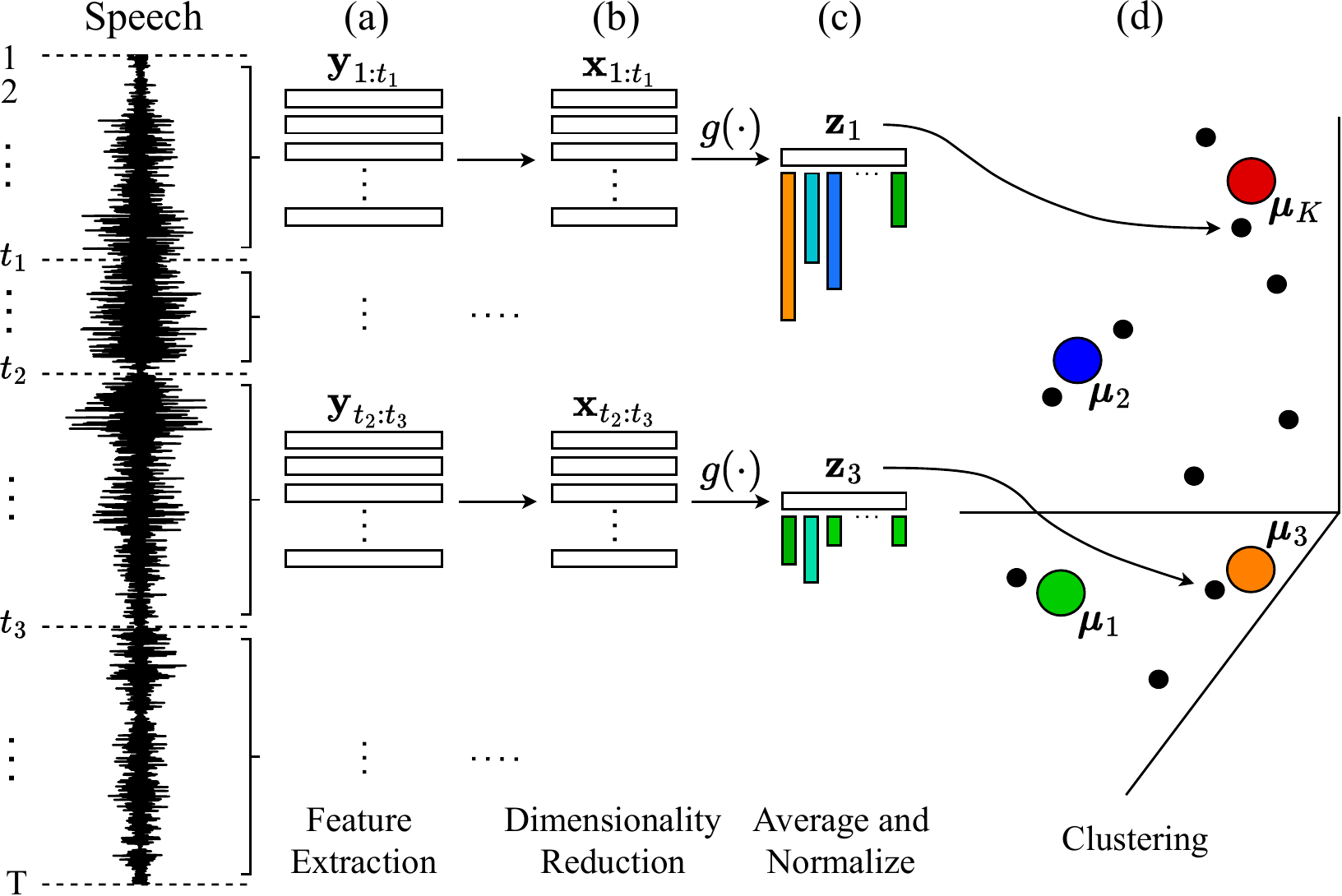}}
    \caption{
    Our lexicon-building step. After extracting frame-level features (a), PCA dimensionality reduction is applied (b). For each segment from the prominence-based approach (Fig.~\ref{fig:tti}), an averaged embedding is obtained~(c). These are $K$-means clustered (d) to get a lexicon.}
    \label{fig:cluster}
\end{figure}

The lexicon building step, illustrated in Fig.~\ref{fig:cluster}, takes the word boundaries (the dashed lines on the left side of the figure) and their corresponding speech utterances as input.
Again, utterances are encoded with an intermediate HuBERT layer, resulting in speech features in a high-dimensional space $\mathbf{y}_{t}\in~\mathbb{R}^{D}$, illustrated at (a) in the figure.
For clustering, working with these high-dimensional features can become 
computationally expensive. 
We therefore 
use principal component analysis (PCA) to reduce the dimensionality of the features,
$\mathbf{x}_{t}\in \mathbb{R}^{M}$ 
with $M < D$,
without loosing much phonetic information~\cite{ramon_awe}.
In this case, the HuBERT features are not normalized and can come from a different layer than the one used for boundary detection, i.e.\ we overload the symbol $\mathbf{y}$.

To cluster the variable duration word segments, we turn to acoustic word embeddings, which map variable-length speech segments to fixed-dimensional vectors~\cite{andrew_awe, keith_awe, kamper_awe}.
We specifically follow the simple 
approach of~\cite{ramon_awe}, where an embedding is obtained by averaging the features in the predicted word segment. 
Formally, a word segment $\mathbf{x}_{t_{1}:t_{2}}$ is transformed into a fixed dimensional embedding vector $\mathbf{z}_{i}=g(\mathbf{x}_{t_{1}:t_{2}})$, where $g$ represents averaging followed by normalization to the unit sphere. 
The result is a set of embeddings $\mathbf{z}_{i} \in \mathbb{R}^M$ (c in Fig.~\ref{fig:cluster}) that are clustered using $K$-means (d).
Our implementation uses the efficient
FAISS\footnote{\url{https://github.com/facebookresearch/faiss}} library for clustering. 
As illustrated on the right side of the figure, each word segment is assigned to the cluster $k$ whose centroid $\bm{\mu}_{k}$ is closest to the segment embedding. 

\section{Revisiting Embedded Segmental K-Means}
\label{sec:eskm}

In this section, we revisit the top-down ES-KMeans method of Kamper et al.~\cite{herman_eskmeans}.
We introduce our own updated version of this method, ES-KMeans+, which incorporates modern features and candidate boundaries.
These updates enable a like-for-like comparison with the simple bottom-up method (above), 
allowing us to isolate the effect of top-down clustering.

\subsection{Embedded Segmental K-Means (ES-KMeans)}

ES-KMeans is a full-coverage word discovery system~\cite{herman_eskmeans}.
It is built on iterative clustering and segmentation steps using dynamic programming.
The clustering step uses $K$-means to cluster the current speech segmentation, while the segmentation step uses the resulting $K$-means model to select boundaries that improve the objective function.

In the original ES-KMeans, utterances are encoded using Mel-frequency cepstral coefficients (MFCCs).
To make the algorithm computationally tractable, boundaries are only allowed at a limited number of candidate positions, provided by a first-pass system.
In the original ES-KMeans, SylSeg~\cite{okko_sylseg} is used to determine a set of candidate boundaries.
This unsupervised signal processing method attempts to find syllable boundaries.
Since ES-KMeans can select an optimal subset of the candidate boundaries, the idea is that some syllable boundaries will be removed, with only word boundaries remaining.

The method starts with a random initialization of the candidate boundaries.
Clustering the current segmentation with $K$-means requires comparing variable-length word segments.
An embedding function $g$ is used to find fixed-length vectors that represent these segments.
Concretely, an arbitrary speech segment $\mathbf{x}_{t_{1}:t_{2}}$, is represented by the embedding $\mathbf{z}_{i}=g(\mathbf{x}_{t_{1}:t_{2}})$, where $\mathbf{z}_{i} \in \mathbb{R}^{M}$.
In~\cite{herman_eskmeans}, $g$ flattens uniformly subsampled features from a segment to get an embedding.

To re-segment, or optimize the segmentation, the current segmentation is “forgotten” and a new segmentation is found.
Candidate boundaries that best match the current $K$-means model are selected using the Viterbi dynamic programming algorithm.
The selection is done by minimizing the cost of an utterance's segmentation, which is calculated as the sum of its segments' embedding scores.

Concretely, an embedding score is defined as:
\begin{equation}\label{eq:eskm_embed_score}
d(\mathbf{z})\triangleq \textrm{len}(\mathbf{z}) \|\mathbf{z}-\bm{\mu}_{\mathbf{z}}^{*}\|^{2},
\end{equation}
where $\textrm{len}(\mathbf{z})$ is the number of frames spanned by the embedding $\mathbf{z}$, and $\bm{\mu}_{\mathbf{z}}^{*}$ is the cluster centroid closest to $\mathbf{z}$.
The length factor is necessary, otherwise the model will always 
favor a trivial solution with the smallest possible word-like segments.
The segmentation objective under a fixed clustering is then
\begin{equation}\label{eq:eskm_cost}
\underset{\mathcal{Q}}{\textrm{min}} \sum_{\mathbf{z}\in \mathcal{Z}(\mathcal{Q})} d(\mathbf{z}),
\end{equation}
where $\mathcal{Z}(\mathcal{Q})$ represents the set of embeddings under a segmentation $\mathcal{Q}$.

\begin{figure}[!t]
    \centerline{\includegraphics[width=\linewidth]{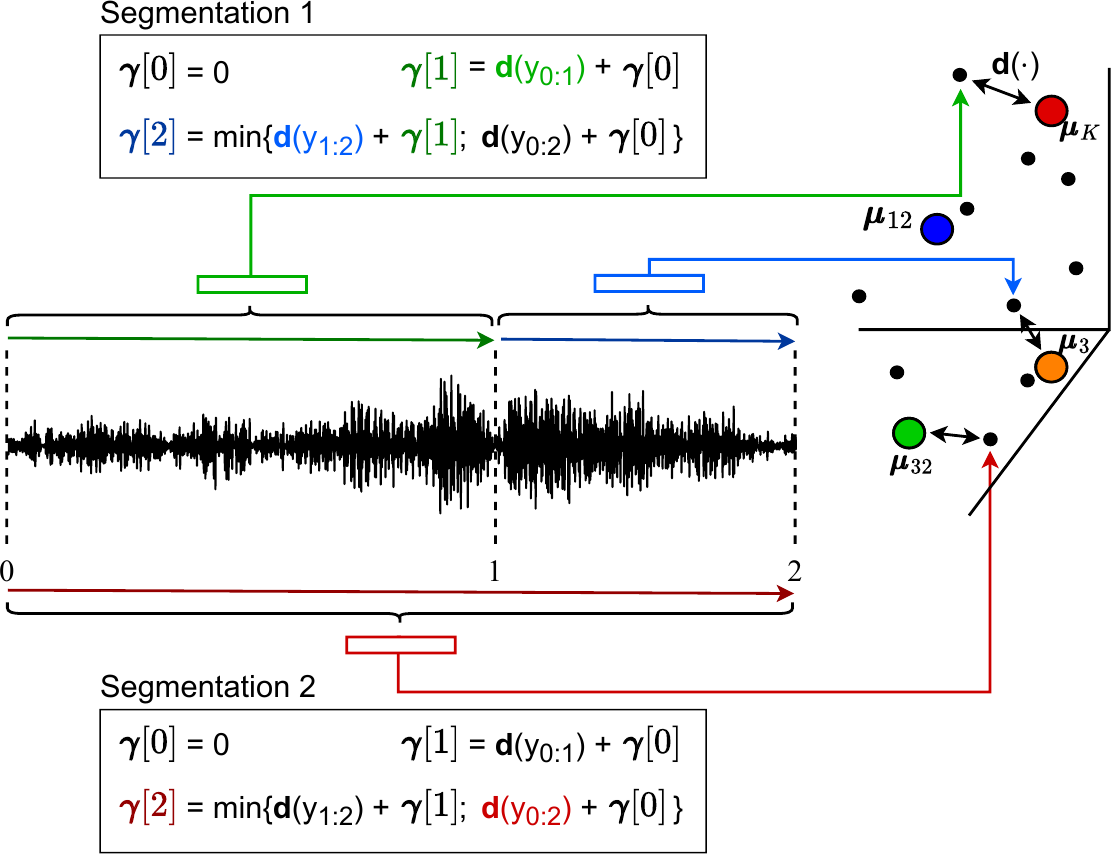}}
    \caption{
    The dynamic programming segmentation step of ES-KMeans. Candidate boundaries are shown with dashed lines. In this example, there are two possible options to segment the utterance. The first uses the intermediate boundary, while the second adds no extra boundaries. The optimal segmentation is determined by the option with the lowest cost.}
    \label{fig:eskm_seg}
\end{figure}

Equation~\ref{eq:eskm_cost} is optimized per utterance using the Viterbi algorithm.
The algorithm uses forward variables that define the minimum score from the beginning of the utterance to the boundary at frame $t$.
Forward variables are calculated as:
\begin{equation}\label{eq:eskm_viterbi}
\gamma[t]=\overset{t}{\underset{l=1}{\textrm{min}}} \{ d(g(\mathbf{y}_{t-l+1:t})) + \gamma[t-1]\}.
\end{equation}
For a particular utterance, we start 
with $\gamma[0]=0$ and then calculate $\gamma[t]$ for all frames $t$ defined as candidate boundaries.
An example is shown in Fig.~\ref{fig:eskm_seg}.
After calculating the forward variables for an utterance ($\gamma[1], \gamma[2]$ in the figure), the optimal segmentation is selected by starting at the last frame $\gamma[T]$ and backtracking through the forward variables, at each step selecting the optimal next step.
The example in Fig.~\ref{fig:eskm_seg} will start at $\gamma[2]$ and either backtrack to the starting boundary via boundary one (Segmentation 1), or skip boundary one and go directly to the starting boundary (Segmentation 2).
The segmentation with the lowest overall cost is selected.

ES-KMeans operates on one utterance at a time: The existing segmentation of an utterance is discarded, and the utterance is re-segmented from scratch.
To update the lexicon, the old segment embeddings for that utterance are removed from the clustering model, and the new embeddings are added.
All other utterances remain unchanged during this process.
This segmentation and clustering step is repeated for each utterance until a final segmentation and lexicon are obtained.

\subsection{Our Updated Method (ES-KMeans+)}
\label{sec:eskm+}

Although ES-KMeans is an older approach, it 
still achieves 
competitive performance~\cite{herman_dpdp_hu}.
To get the most out of this 
method, we revisit the algorithm and update it using modern approaches.
Concretely, we replace the SylSeg~\cite{okko_sylseg} boundary constraints with the prominence-based approach~\cite{ankita_tti} described in Section~\ref{sec:psc}. 
We replace MFCCs with HuBERT features and follow the same clustering approach as in Fig.~\ref{fig:cluster}: PCA-projected features are averaged, normalized, and clustered using $K$-means.
Instead of the per-utterance nature of the original algorithm, we operate in batches, entirely re-segmenting all utterances and only then clustering the new hypothesized segments of each utterance.
This allows us to again use the efficient FAISS library for clustering.
The benefits of each of these changes were verified in developmental experiments, leading to a much improved and scalable implementation. 

After these updates, ES-KMeans+ and the bottom-up approach of Section~\ref{sec:psc} are very similar. The latter corresponds to the first iteration of ES-KMeans+, just before word segmentation is performed. The bottom-up approach is therefore much faster (no further refinements), but ES-KMeans+ has the benefit that it can decide to remove prominence-detected boundaries if these are poorly matched to the clustering model.

\section{Experimental Setup}
\label{sec:exp_setup}

\begin{table*}[!t]
    \mytable
    \caption{Performance (\%) of Our Two Approaches
    and Other State-of-the-Art Methods\\for Word Segmentation ($R$-val and $F_1$) and Lexicon Building (NED) on Track 2 of the ZeroSpeech Challenge.}
    \footnotesize
    \begin{tabularx}{\linewidth}{@{}l@{\ \ }C@{\ \ }C@{\ \ }CC@{\ \ }C@{\ \ }CC@{\ \ }C@{\ \ }CC@{\ \ }C@{\ \ }CC@{\ \ }C@{\ \ }C@{}}    
        \toprule
        & \multicolumn{3}{c}{English} & \multicolumn{3}{c}{French} & \multicolumn{3}{c}{Mandarin} & \multicolumn{3}{c}{German} & \multicolumn{3}{c}{Wolof}\\
        \cmidrule{2-4}
        \cmidrule(l){5-7}
        \cmidrule(l){8-10}
        \cmidrule(l){11-13}
        \cmidrule(l){14-16}
        Model & NED & $R$-val. & Token $F_{1}$ & NED & $R$-val. & Token $F_{1}$ & NED & $R$-val. & Token $F_{1}$ & NED & $R$-val. & Token $F_{1}$ & NED & $R$-val. & Token $F_{1}$\\
        \midrule
        VG-HuBERT~\cite{vg_hubert} & 41.0 & 59.8 & \textbf{24.0} & 62.0 & 44.0 & 15.0 & 73.0 & 32.5 & 19.0 & 56.0 & 21.9 & \textbf{15.0} & 92.0 & 59.7 & 9.0 \\
        Self-expressing AE~\cite{bhati_se_ae} & 89.5 & -61.4 & 6.6 & 89.0 & -66.7 & 6.3 & 96.6 & -32.5 & 12.1 & 89.6 & -116.8 & 6.3 & 82.7 & -27.8 & 12.6 \\
        DPDP~\cite{herman_dpdp_hu} & 41.7 & \textbf{63.2} & 19.6 & 66.0 & 60.3 & 11.6 & 86.0 & \textbf{67.9} & 24.5 & 56.8 & \textbf{49.2} & 12.5 & 72.2 & 66.9 & 13.1 \\ 
        ES-KMeans~\cite{herman_eskmeans} & 73.2 & 51.5 & 19.2 & 68.7 & 37.2 & 6.3 & 88.1 & 23.3 & 8.1 & 66.2 & 17.1 & 11.5 & 72.4 & 58.3 & 10.9 \\
        ES-KMeans+~[ours]& 33.5 & 50.0 & 14.7 & \textbf{43.2} & 56.3 & \textbf{20.0} & \textbf{65.5} & 56.9 & \textbf{25.1} & \textbf{42.8} & 25.5 & 9.7 & \textbf{56.2} & \textbf{69.4} & \textbf{24.5} \\
        Prom. Seg. Clus.~[ours] & \textbf{32.9} & 60.6 & \textbf{24.0} & 47.9 & \textbf{61.0} & 17.2 & 71.4 & 58.2 & 22.7 & 44.3 & 41.8 & 10.9 & 59.3 & 67.9 & 19.5 \\
        \bottomrule
    \end{tabularx}
    \label{tbl:zrc_results}
\end{table*}

\subsection{Data}

We perform our main experiments on Track 2 of the ZeroSpeech Challenge~\cite{ewac_zrc}, which has become an established benchmark for comparing word discovery methods. The evaluation
covers
five languages: English, French, Mandarin, German, and Wolof, respectively consisting of 45, 24, 2.5, 25, and 10 hours of speech. 
The challenge encourages participants to develop methods that can be applied to any language and is one of the most comprehensive benchmarks for unsupervised speech systems.
For development experiments and further analysis, we additionally use
the dev-clean subset of LibriSpeech~\cite{librispeech}, 
with 5.4 hours of English speech from 40 speakers.

\subsection{Evaluation Metrics}

In addition to word boundary precision, recall, and $F_1$, word segmentation is evaluated using $R$-value and token $F_{1}$.
For all of these metrics, higher is better.
$R$-value measures how close hypothesized word boundaries are to an 
ideal operating point with a
100\% recall and 0\% over-segmentation~\cite{okko_rval}.
Here, over-segmentation is the ratio of the number of hypothesized boundaries relative to the number of ground-truth boundaries~\cite{petek_os}.
Token $F_{1}$ evaluates how well the hypothesized word tokens match ground-truth tokens, requiring both predicted boundaries to be correct to receive credit.
In our evaluations, these metrics are calculated with a boundary tolerance of 20 ms.
Therefore, a hypothesized boundary can lie within 20 ms to either side of a ground-truth boundary to be considered a hit.

The quality of a lexicon is evaluated using normalized edit distance (NED)~\cite{ned}.
This relies on phonemic transcriptions found by forced alignments. 
Discovered word tokens are mapped to their overlapping phoneme sequence, and the NED is calculated between all phoneme sequences of the segments within each cluster. Lower NED is better.
NED is considered together with bitrate, which determines the average number of bits per second (bits/s) required to transmit an encoded sequence.
Assuming a constant NED, a lower bitrate is better since less data is required to convey the same information.

\subsection{System Implementation Details}

For both the prominence-based word segmentation step of Section~\ref{sec:psc} and for the potential boundary set of ES-KMeans+ in Section~\ref{sec:eskm+}, we extract features from the 9th HuBERT layer, based on the findings of~\cite{ankita_tti}.
We use the fine-tuned HuBERT model from~\cite{van_Niekerk_hu_soft} as our content encoder, based on its performance in preliminary experiments (not reported here).
The smoothing window controls how emphasized the dissimilarity curve is, while the prominence threshold determines which peaks are chosen as word boundaries.
We set these parameters 
by finding a 
balance between 
NED and $R$-value on development data:
For our bottom-up word discovery approach, we select a four-frame window 
with a 0.75 prominence threshold.
For ES-KMeans+,
we opt for a five-frame window with a threshold of 0.3. 
This high-recall hyperparameter setting for ES-KMeans+ creates
more word boundaries than are needed, enabling the method to choose the best subset of these candidate boundaries.

When clustering, both in our bottom-up method and ES-KMeans+, we extract features from the 12th fine-tuned HuBERT~\cite{van_Niekerk_hu_soft} layer and reduce the feature dimensionality to 250 dimensions using PCA, as illustrated at (b) in Fig.~\ref{fig:cluster}.
These settings are based on development experiments.

Other ES-KMeans+ parameters are kept the same as in the original ES-KMeans paper~\cite{herman_eskmeans}: The segmentation is randomly initialized at half of the candidate boundaries, each
segment has a minimum duration of five frames, and a segment can at most span four candidate boundaries.
We run ES-KMeans+ for five iterations.

To enable a fair comparison to previous work,
we use the same number of $K$-means clusters as in the original ES-KMeans~\cite{herman_eskmeans} and DPDP~\cite{herman_dpdp_hu} papers: For the ZeroSpeech challenge we use 43k, 29k, 3k, 29k, and 3.5k clusters for English, French, Mandarin, German, and Wolof.
These numbers were set to 10\% of the number of tokens discovered by SylSeg~\cite{okko_sylseg}.
For experiments on 
LibriSpeech dev-clean, we use 14k clusters.

We compare our systems with representative bottom-up and top-down methods from Section~\ref{sec:background}.
We also compare to the visually-grounded HuBERT (VG-HuBERT) method~\cite{vg_hubert}. 
This approach relies on paired instances of unlabeled speech with an image of a scene---although unsupervised, it should be considered as distinct given the use of an additional modality.

\section{Experiments}

We begin 
by comparing our methods to previous full-coverage unsupervised segmentation and clustering systems.
This is done with a specific focus on comparing our two approaches within the context of other bottom-up and top-down word discovery methods.
After comparing the efficiency of ES-KMeans+ with the purely bottom-up approach, we assess the impact of the design choices within our updates to ES-KMeans.
Finally, we study the inner workings of our methods to compare the two methodologies:
We examine the types of boundaries they find and how these choices influence their resulting lexicons.
Finally, we analyze the role of the candidate boundaries and how ES-KMeans+ selects among them.

\subsection{Unsupervised Word Discovery}

\begin{table}[!t]
    \mytable
    \caption{Runtime (min) of Prominence Segmentation with Clustering\\and ES-KMeans+.}
    \footnotesize
    \begin{tabularx}{\linewidth}{@{}l@{}CCCCC@{}}
        \toprule
        Model & English & French & Mandarin & German & Wolof\\
        \midrule
        ES-KMeans+~[ours] & 765 & 330 & 5 & 316 & 8 \\ 
        Prom. Seg. Clus.~[ours] & \textbf{146} & \textbf{68} & \textbf{1} & \textbf{50} & \textbf{3} \\
        \bottomrule
    \end{tabularx}
    \label{tbl:runtime}
\end{table}

Word discovery performance on the ZeroSpeech benchmark is presented in Table~\ref{tbl:zrc_results}.
When comparing 
the prominence-based method in the last line to 
ES-KMeans+,
we see
a tradeoff between
lexicon quality (NED) and word segmentation performance ($R$-value):
The bottom-up method achieves better $R$-value on all languages except Wolof, while ES-KMeans+ gives better NED in all cases but English.
On English, prominence segmentation with clustering outperforms ES-KMeans+ across all metrics.

Compared to the other approaches, 
both our systems achieve several 
state-of-the-art results and improve upon the original ES-KMeans. 
Prominence segmentation achieves a good tradeoff between metrics,
improving on all scores compared to 
the bottom-up self-expressing auto encoder (AE)
~\cite{bhati_se_ae}.
While our bottom-up method does not achieve as high $R$-values as the bottom-up DPDP approach~\cite{herman_dpdp_hu}, it achieves better $R$-values than ES-KMeans+ on all languages except for Wolof.

In some cases, token $F_1$ is high but $R$-value is low.
This represents cases where a system receives credit for correct boundaries,
but this is because it over-segments, proposing many more boundaries than there actually are.
VG-HuBERT~\cite{vg_hubert} on German is one of such cases.
Neither of our systems shows this effect,
indicating that we discover actual words 
rather than artificially achieving high scores by
over-segmenting the utterances.

One
limitation of the prominence-based method is its 
dependence on the quality of the word segmentation step 
since it cannot remove bad boundaries like DPDP or ES-KMeans.
Due to its iterative refinement steps, ES-KMeans+ is significantly slower than its bottom-up counterpart which only segments and clusters once.
We quantify this difference by measuring the runtime of each method, reported in Table~\ref{tbl:runtime}.
While
ES-KMeans+ performs marginally better than the prominence-based
approach on several metrics, Table~\ref{tbl:runtime} shows that ES-KMeans+ is four to five times slower.

Because of the tradeoff between $R$-value, NED, and runtime, there is no clear choice between the purely bottom-up prominence segmentation approach and the top-down cluster-informed ES-KMeans+ approach. 
Next, we show the effects of our updates to ES-KMeans, and then we analyze this tradeoff to make the comparison more definitive.

\subsection{Effect of our Design Choices}

\begin{table}[!t]
    \mytable
    \caption{Ablation Study of the Performance (\%) of the Main Components of ES-KMeans(+) on LibriSpeech dev-clean. A Boundary Tolerance of 20 ms is Allowed.}
    \footnotesize
    \begin{tabularx}{\linewidth}{@{}l@{\ \ \ }l@{\ \ \ }C@{\ \ \ }C@{\ \ \ }C@{}}
        \toprule
        & Candidate boundaries & NED & $R$-value & Token $F_{1}$ \\
        \midrule
        & \underline{\textit{MFCC-based lexicon}} & & & \\
        \ii{1} & SylSeg (ES-KMeans~\cite{herman_eskmeans}) & 80.4 & 41.1 & 5.8 \\
        \ii{2} & Prom. Seg. on MFCCs & 83.2 & 40.1 & 3.9 \\
        \vspace{0.25cm}
        \ii{3} & Prom. Seg. on HuBERT & 80.1 & 52.0 & 14.8 \\
        & \underline{\textit{HuBERT-based lexicon}} & & & \\
        \ii{4} & SylSeg & 46.0 & 41.5 & 7.2 \\
        \ii{5} & Prom. Seg. on MFCCs & 53.5 & 41.5 & 6.0 \\
        \ii{6} & Prom. Seg. on HuBERT (ES-KMeans+ [ours]) & \textbf{42.5} & \textbf{53.2} & \textbf{19.5} \\
        \bottomrule
    \end{tabularx}
    \label{tbl:ablation}
\end{table}

ES-KMeans+ gives competitive performance by utilizing self-supervised features for both clustering and candidate boundary selection (second to last line, Table~\ref{tbl:zrc_results}).
To support this argument, we look at how performance is impacted when the components of ES-KMeans+ rely on SylSeg~\cite{okko_sylseg} and MFCCs rather than self-supervised features.

Table~\ref{tbl:ablation} is split into two sections based on the representations used to build a lexicon.
Like in the original ES-KMeans, the first section builds a lexicon using subsampled MFCCs while the second section uses averaged HuBERT features as described in Section~\ref{sec:psc}.
Both of these sections pair their lexicon-building step with three different sets of candidate boundaries.

To isolate the influence of the candidate boundaries, we keep the lexicon features
fixed and consider 
each section of Table~\ref{tbl:ablation} on its own. 
In the 
MFCC 
section (top), line~1 uses SylSeg boundaries, resulting in a system equivalent to the original ES-KMeans.
Comparable segmentation performance ($R$ and $F_1$) is observed in line~2, where SylSeg is replaced with the prominence-based approach from~\cite{ankita_tti} but applied on MFCCs instead of HuBERT.
Only in line~3---where HuBERT features are used for prominence-based segmentation---does ES-KMeans+ show an improvement in both $R$-value and token $F_1$.
These improvements suggest that HuBERT features capture spectral variations in speech more effectively than MFCCs, and that SylSeg is limited by its signal processing-based approach.
These findings are corroborated in the bottom section of the table, where HuBERT features are used for lexicon learning.

We now turn to the question of which features are better for lexicon learning. 
We do this by comparing systems with the same candidate boundaries.
Using the best-performing boundaries---prominence-based segmentation with HuBERT---the comparison between lines~3 and~6 in Table~\ref{tbl:ablation} shows that modern features are much better for lexicon building:
The NED improves by 37.6\% absolute when using averaged HuBERT features compared to subsampled MFCCs.

Comparing our final ES-KMeans+ system in line~6 to the original ES-KMeans in line~1,
we see that the combination of improved boundaries and improved features is what leads to the competitive performance of the 
updated method. But both
of these improvements can actually 
be attributed to the representation capabilities of self-supervised speech models.

Having confirmed that ES-KMeans+ is an
improvement on the original method,
the remainder of this paper asks whether the higher computational cost of the top-down ES-KMeans+ is worth the marginal gain in performance over the simple bottom-up method (as shown in Table~\ref{tbl:runtime}).
To answer this question, we start by looking at the differences in the discovered boundaries.

\subsection{What Types of Boundaries are Discovered?}

\begin{table}[t!]
    \mytable
    \caption{Word- and Syllable-Level Evaluation of Our Methods on LibriSpeech dev-clean. A Boundary Tolerance\\of 20 ms is Allowed.}
    \footnotesize
    \begin{tabularx}{\linewidth}{@{}l@{\ \ \ }C@{}C@{}C@{}C@{}c@{}}
        \toprule
        & \multicolumn{3}{c}{Word/Syllable boundary} & \multicolumn{1}{c}{Token} & \\
        \cmidrule(lr){2-4}
        \cmidrule(lr){5-5}
        Model & Prec. & Rec. & $F_{1}$ & $F_{1}$ & $R$-value \\
        \midrule
        \underline{\textit{Word level}} & & & & & \\
        ES-KMeans+~[ours] & 45.3 & 49.0 & 47.1 & 19.5 & 53.2 \\ 
        \vspace{0.25cm}
        Prom. Seg. Clus.~[ours] & 42.2 & 42.5 & 42.4 & 15.6 & 50.7 \\
        \underline{\textit{Syllable level}} & & & & & \\
        ES-KMeans+~[ours] & 52.3 & 43.5 & 47.5 & 20.4 & 56.5 \\ 
        Prom. Seg. Clus.~[ours] & 41.4 & 32.0 & 36.1 & 11.7 & 48.1 \\
        \bottomrule
    \end{tabularx}
    \label{tbl:syll_eval}
\end{table}

Although we have assessed 
our systems in terms of word discovery performance, we have not looked at what types of boundaries
they actually predict.
Informally listening to the
segmented units 
reveals that both methods often discover monosyllabic words.
To quantitatively investigate this, we compare word- and syllable-level boundary evaluations.

Ground-truth syllable boundaries are not provided with LibriSpeech. 
Therefore, syllable boundaries are found using a rule-based system\footnote{\url{https://github.com/kylebgorman/syllabify}} that converts ground-truth phonemic transcriptions to syllable alignments.
These syllable boundaries are not 
perfect, as in some cases the rule-based system merges phones that span across words, resulting in skipped word boundaries.
But it is sufficient for our comparison in Table~\ref{tbl:syll_eval}.

On LibriSpeech dev-clean, ES-KMeans+ finds better word and syllable boundaries than prominence segmentation with clustering.
This improvement is especially pronounced in the syllable-level evaluation (bottom half of Table~\ref{tbl:syll_eval}), where ES-KMeans+ achieves an 8.7$\%$ absolute gain in syllable token $F_1$ over the bottom-up method.

Comparing ES-KMeans+ across the word- and syllable-level evaluations, seen in the top and bottom halves of Table~\ref{tbl:syll_eval}, the method performs better on the syllable discovery task.
The token $F_1$ scores of ES-KMeans+ are similar for both the word and syllable evaluations. 
This suggests that many of the words contributing to its word-level performance are likely monosyllabic.
In contrast, prominence segmentation with clustering detects fewer syllable boundaries than word boundaries, as its performance improves across all metrics from syllable- to word-level evaluations.
It is worth reiterating that although both of our methods use prominence-based boundaries, their hyperparameter settings differ: ES-KMeans+ uses a higher-recall detector, while the bottom-up approach is tuned directly on development data for the task of word discovery.

Therefore, ES-KMeans+ tends to find segments closer to syllables, while our bottom-up method finds coarser segments corresponding to words spanning multiple syllables.
In general, on the word segmentation task, the top-down methodology improves on the simple bottom-up system.

\subsection{What Types of Units are Discovered?}
\label{sec:units}

\begin{figure}[!t]
     \centering
     \subfloat[]{\includegraphics[width=\linewidth]{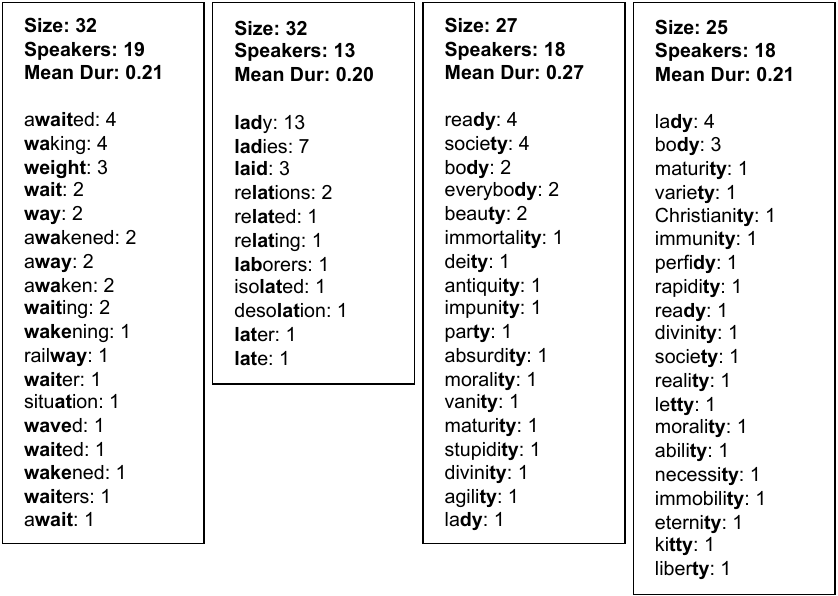}
     \label{fig:clust_analysis_eskm}}
     \vspace{1.5mm} 
     \subfloat[]{\includegraphics[width=\linewidth]{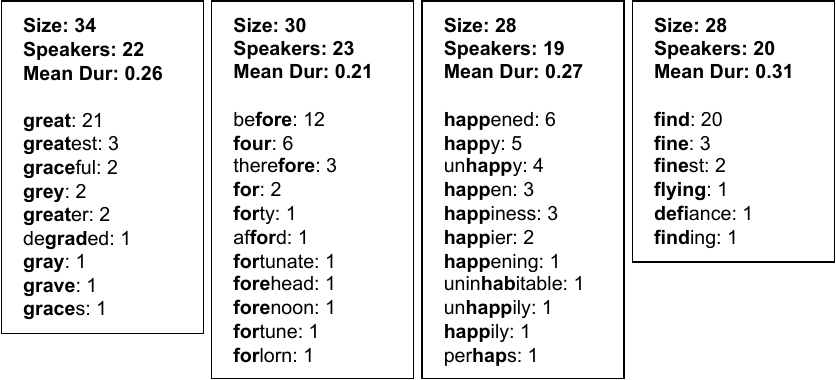}
     \label{fig:clust_analysis_psc}}
     \caption{Transcription of word units present in the largest clusters of 
     (a) ES-KMeans+ and (b) prominence segmentation with clustering. The size in terms of tokens, the number of unique speakers, and the mean duration of the audio segments in seconds is provided for each cluster.}
     \label{fig:clust_analysis}
\end{figure}

Given that ES-KMeans+ discovers finer-grained boundaries than prominence segmentation with clustering, 
how do the discovered units differ?
Concretely, how are the segments clustered?
And how does the clustering interact with the predicted segmentation?

We specifically look at some of the largest clusters of both our methods.
For each discovered word token, we determine the ground-truth label with which it has the greatest overlap.
Then, the total number of tokens, the number of unique speakers, and the mean duration of the tokens in seconds are determined for each cluster.
Fig.~\ref{fig:clust_analysis} visualizes the content of the clusters by listing the number of occurrences of their word-like types.
The segment of the ground-truth transcription actually contained in the discovered word segment is shown in boldface. 

The mean token durations of ES-KMeans+ in Fig.~\ref{fig:clust_analysis_eskm} (0.21 seconds for the leftmost cluster) and prominence segmentation with clustering in Fig.~\ref{fig:clust_analysis_psc} (0.26 seconds for the leftmost cluster) strengthen our notion that ES-KMeans+ finds shorter segments than the bottom-up method.

The boldface segments of ES-KMeans+ are sometimes monosyllabic words like “wait”, but in other cases are common sub-word units like “dy/ty”, as in the two rightmost clusters in Fig.~\ref{fig:clust_analysis_eskm}.
Here, over-clustering occurs where 
multiple clusters end up containing the same type.
Although the bottom-up method's types are usually a complete word, e.g., ``great'', ``four/for'', and ``find'', they are commonly segments of larger words e.g., ``greatest'' and ``before''.

When the beginning of a word like ``waited'' is placed in a ``wait'' cluster, where does the rest of the word end up?
To answer this question, we investigated where
remainders 
(non-boldface letters) are clustered.
We do not show the analysis here, but briefly summarize the findings. As an example, for the ``dy/ty''
cluster in Fig.~\ref{fig:clust_analysis_eskm} with size~27, we confirmed 
that the ``moral'' segment in the instance of ``morality'' is clustered with other ``moral'' instances.
For prominence segmentation with clustering in Fig.~\ref{fig:clust_analysis_psc}, this also happens.
However, the fixed segmentation in the bottom-up approach leads to cases where a word like ``before'' in the cluster with size~30 is segmented in one utterance as ``them-be'' and ``fore''.
Here, ``them-be'' is clustered with other instances of ``be''.
ES-KMeans+ has the advantage that it can select boundaries to fix these kinds of occurrences.

Both methods show promising signs of disregarding speaker-specific information, as the clusters in Fig.~\ref{fig:clust_analysis} contain many speakers.
However, in both cases there is a large degree of over-clustering.

\subsection{Limitations of Candidate Boundaries}
\label{sec:cand_bound}

\begin{table}[!t]
    \mytable
    \caption{Performance (\%) of Our Methods with Different Candidate Boundaries on LibriSpeech dev-clean. A Boundary Tolerance\\of 20 ms is Allowed.}
    \footnotesize
    \begin{tabularx}{\linewidth}{@{}l@{\ \ \ }lc@{\ \ }c@{\ \ }c@{\ \ }c@{}c@{\ \ }c@{}@{}}
        \toprule
        & & \multicolumn{3}{c}{Word boundary} & \multicolumn{1}{c}{Token} & & \\
        \cmidrule(lr){3-5}
        \cmidrule(lr){6-6}
        & Candidate boundaries & Prec. & Rec. & $F_{1}$ & $F_{1}$ & $R$-value & NED \\
        \midrule
        & \underline{\textit{ES-KMeans+~[ours]}} & & & & & & \\
        \ii{1} & Prom. Seg. & 45.2 & 49.0 & 47.1 & 19.5 & 53.2 & 42.3 \\
        \ii{2} & $\textrm{GT}_{\textrm{phones}}$ & 32.8 & 81.7 & 46.8 & 12.8 & -34.4 & 23.5 \\
        \ii{3} & $\textrm{GT}_{\textrm{syllables}}$ & 59.4 & 72.4 & 65.3 & 40.1 & 65.0 & 26.0 \\
        \ii{4} & $\textrm{GT}_{\textrm{words}}$ & 100.0 & 94.8 & 97.3 & 93.7 & 91.4 & 30.0 \\
        \ii{5} & $\textrm{GT}_{\textrm{words}}$ + Prom. Seg. & 62.1 & 85.6 & 72.0 & 43.1 & 61.3 & 36.4 \\
        \vspace{0.25cm}
        \ii{6} & Max Rec. Prom. Seg. & 33.5 & 66.0 & 44.5 & 15.0 & 2.3 & 41.3 \\
        & \underline{\textit{Prom. Seg. Clus.~[ours]}} & & & & & & \\
        \ii{7} & Prom. Seg. & 42.3 & 51.9 & 46.6 & 19.0 & 48.5 & 44.8 \\
        \ii{8} & $\textrm{GT}_{\textrm{phones}}$ & 30.8 & 100 & 47.1 & 11.0 & -92.1 & 20.4 \\
        \ii{9} & $\textrm{GT}_{\textrm{syllables}}$ & 59.6 & 77.5 & 67.4 & 43.9 & 62.7 & 26.3 \\
        \ii{10} & $\textrm{GT}_{\textrm{words}}$ & 100 & 100 & 100 & 100 & 100 & 31.1 \\
        \ii{11} & $\textrm{GT}_{\textrm{words}}$ + Prom. Seg. & 55.5 & 100 & 71.4 & 39.3 & 31.6 & 39.8 \\
        \ii{12} & Max Rec. Prom. Seg. & 29.6 & 72.4 & 42.0 & 12.9 & -34.5 & 47.4 \\
        \bottomrule
    \end{tabularx}
    \label{tbl:eskm_gt_bound}
\end{table}

So far, all candidate boundaries have come from the prominence-based method of~\cite{ankita_tti}.
What is the upper bound of ES-KMeans+ performance with more ideal boundaries? How much is the approach constrained by the non-perfect candidate boundaries?

Table~\ref{tbl:eskm_gt_bound} is divided into two sections.
The top part 
shows the final ES-KMeans+ scores for the different candidate boundary systems.
The bottom part, 
where the 
prominence segmentation with clustering method is used, is a baseline showing where the candidate boundaries start, before ES-KMeans+ sub-selects boundaries.
The difference in the scores of the two sections illustrates how ES-KMeans+ (top) drifts from its initial boundaries (bottom).
This drift reveals how boundaries are selected and whether the top-down method does anything useful.

We look at six different candidate boundary settings.
The first is the prominence-based boundaries (Prom. Seg.) used so far.
Ground-truth (GT) phone, syllable, and word boundaries are used to simulate 
perfect candidate boundary selections with different granularities. 
To see if ES-KMeans+ can select the true word boundaries from a set, we add our prominence-based boundaries to the ground-truth words set ($\textrm{GT}_{\textrm{words}}$ + Prom. Seg.).
Lastly, we use the prominence-based method with a smoothing window size of three frames and a prominence threshold of $\textrm{1}\cdot\textrm{10}^{-\textrm{4}}$ (Max Rec. Prom. Seg.), instead of the settings in Section~\ref{sec:exp_setup}---this gives many more boundaries, maximizing
the recall of the boundary set, but also increasing
over-segmentation.
All systems are initialized with boundaries at every possible candidate position.

The most striking feature of Table~\ref{tbl:eskm_gt_bound} is how ES-KMeans+ (top) trades word boundary recall in favor of precision as it moves away from its initialization (bottom).
This can be seen in 
going from line~11 to line~5 or from line~12 to line~6.
With this shift, ES-KMeans+ consistently improves the $R$-value of the candidate boundaries, without 
sacrificing NED. 

Although the fine-grained ground-truth phone boundaries in line~2 over-segments (low $R$-value),
ES-KMeans+ actually improves the boundary scores relative to its 
initialization in line~8.
When adding many prominence-based boundaries to the candidate set in line~12, ES-KMeans+ improves all scores except recall, increasing the $R$-value by 36.8$\%$ absolute in line~6.
Here, we start to see what ES-KMeans+ does well.

What makes ES-KMeans+ useful is its ability to improve candidate boundaries that over-segment without worsening the other metrics.
The system where prominence-based boundaries are added to the ground-truth word boundaries in line~5 ends up with better scores than the 
original system in line~1, although starting with a much worse $R$-value (line~11 compared to line~7).
This shows that for ES-KMeans+ the candidate boundaries can be slightly over-segmented, 
as long as there are also correct word 
boundaries contained in the candidate set.

Given the perfect word boundaries in line~10, the far from perfect NED of 31.1\% highlights that our
lexicon-building step is far from perfect.
Although all word boundary scores are worsened by ES-KMeans+ in line~4, the NED is similar (going from 31\% to 30\%). 
If the acoustic word embeddings were truly representative of 
the
word segments and these were perfectly clustered, the NED moving from line~10 to line~4 should have worsened. 
This comparison also shows that, given perfect boundaries, ES-KMeans+ moves away from the truth since all it can do is merge boundaries, resulting in a worse system.

\begin{figure}
    \centerline{\includegraphics[width=\linewidth]{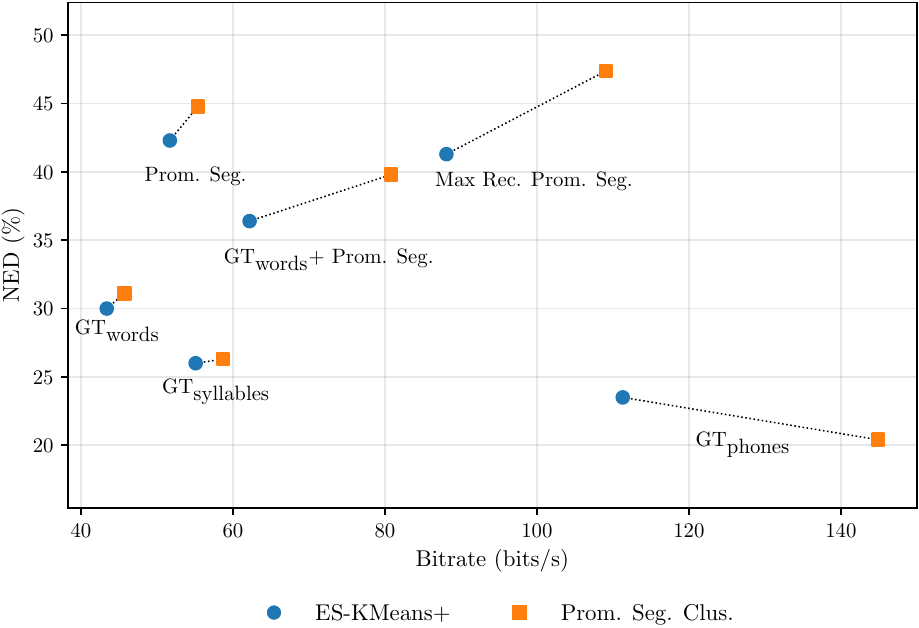}}
    \caption{Lexicon quality of ES-KMeans+ (blue dot) and prominence segmentation with clustering (orange square) initialized on different candidate boundary sets on LibriSpeech dev-clean. A perfect system has a NED (\%) of zero with a low bitrate (bits/s).}
    \label{fig:bitrate_ned}
\end{figure}

To further evaluate how
the subsequent refinement steps of
ES-KMeans+ influences 
lexicon learning,
we investigate the change in lexicon quality when moving from the initialization to the ES-KMeans+ output.
Lexicon quality (in terms of NED) cannot be considered in isolation. 
The lexicon influences how compactly the data can be encoded, and this should be taken into account. 
We therefore also measure the bitrate of the resulting encoding~\cite{zrc2019}.
Fig.~\ref{fig:bitrate_ned}
shows the NED and bitrate of the systems with different candidate boundaries.
A perfect system would be in the bottom-left corner of the figure (perfect NED with very few bits).
The line going from the orange {square} (prominence segmentation with clustering) 
to the blue dot (ES-KMeans+) 
shows the drift of the ES-KMeans+ sub-selection strategy as it selects boundaries from the provided candidates.
ES-KMeans+ moves all systems closer to an optimal point of operation, except in 
cases where ground-truth phone or syllable boundaries are used.
Both the NED and bitrate of the four remaining systems (above 30\% on the y-axis) decrease when applying ES-KMeans+ to the candidate set, showing an overall improvement in the lexicon quality.
In these systems, the improvement in NED becomes larger as the improvement in bitrate increases.
Therefore, systems that over-segment the most, like the maximum recall setting, benefit more 
from
ES-KMeans+ refinement.

The only baseline system that ES-KMeans+ worsens in terms of NED is the one using ground-truth phone boundaries (bottom-right corner of Fig.~\ref{fig:bitrate_ned}).
This is to be expected as the low number of phone types in the candidate set is easy to cluster (orange square in the bottom-right corner).
When ES-KMeans+ removes some of these candidate boundaries (blue dot), the number of types increases, making the clustering task more difficult.

In terms of lexicon building, ES-KMeans+ is justified in all of our systems that attempt to find word-like units.
However, if the candidate boundaries are already close to perfect (like the ground-truth word and syllable systems), ES-KMeans+ has little effect on the NED and bitrate. 

\section{Summary and Conclusion}

This paper investigated how full-coverage unsupervised word discovery is impacted when top-down information is incorporated to select word boundaries.
Top-down methods use higher-level information to guide boundary selection, whereas bottom-up methods fix their boundaries before constructing a lexicon by clustering the resulting segments.

We compared two structurally similar methods that differ only in their use of top-down information.
Our
bottom-up prominence-based method predicts word boundaries at peaks in the distances between adjacent self-supervised features and clusters the resulting units using $K$-means.
Our
top-down method is a revisited implementation of the dynamic-programming ES-KMeans method~\cite{herman_eskmeans}:
It iteratively refines a candidate segmentation based on how well the discovered units match the clustering model.

On the five-language ZeroSpeech benchmarks, the two methods
achieve similar state-of-the-art performance.
In general, ES-KMeans+ produces a slightly better lexicon, 
while the bottom-up method is significantly faster.
By examining the predicted boundaries and their impact on lexicon building, we found that ES-KMeans+ tends to identify shorter sub-word segments, often corresponding to syllables rather than full words.
These shorter segments lead to over-clustering, where multiple clusters correspond to the same sub-word type.
On the other hand, the bottom-up method finds segments closer to words.
But these are mostly short monosyllabic words.
This highlights the inability of our word segmentation step to find full or longer words.

Further analyses showed
that candidate boundaries play a major role in the final segmentation of ES-KMeans+, which trades recall for precision as it moves from 
its initial segmentation.
This motivates a specific use case for ES-KMeans+: It is particularly useful
when candidate boundaries require refinement, especially if the candidates slightly over-segment compared to true words.
When candidate boundaries are already well-defined, ES-KMeans+ drifts from these boundaries, and it is better to use a simple bottom-up method.
In many practical cases, a simple bottom-up method is sufficient---with the added benefit of being much faster.

We showed that both of our methods are limited by the lexicon-building step, which relies on $K$-means clustering on averaged HuBERT features~\cite{wei_hubert}.
This was most evident in experiments where the lexicon was still far from perfect, even when replacing the candidate boundaries with ground-truth word boundaries.
A specific problem was over-clustering, where a true word is separated into different clusters.
To address this in future work, three avenues should be considered.
(1) Speech representations must be improved so that different instances of the same word have similar embeddings.
(2) Since $K$-means over-clusters, other clustering methods should be reconsidered.
(3)~For top-down methods, the initial set of boundaries is crucial.
A candidate set must be configured for the method's strengths. If the candidate boundaries are already well-defined, it may be good enough to use a bottom-up method. 

\bibliography{IEEEabrv,mybib}

\end{document}